\documentclass[prb,superscriptaddress,twocolumn,floatfix]{revtex4-1}

\usepackage{graphicx}
\usepackage{amsmath}

\def\on-line-cite#1{[\onlinecite{#1}]}
\def\textdegree{$^{\circ}$}

\begin{document}

\title{{\it Ab initio} $^{27}Al$ NMR chemical shifts and quadrupolar
parameters\\for $Al_2O_3$ phases and their precursors}

\author{Ary R. Ferreira}
\affiliation{ Universidade Federal de Juiz de Fora (UFJF), 
Department of Chemistry, Juiz de Fora, MG, 36036-330, Brazil}

\author{Emine K\"{u}\c{c}\"{u}kbenli}
\affiliation{Scuola Internazionale Superiore di Studi Avanzati (SISSA),
Via Bonomea 265, I-34136 Trieste, Italy}
\affiliation{CNR-IOM DEMOCRITOS Simulation Center, Via Bonomea 265, I-34136 Trieste, Italy}

\author{ Alexandre A. Leit\~{a}o}
\affiliation{ Universidade Federal de Juiz de Fora (UFJF), 
Department of Chemistry, Juiz de Fora, MG, 36036-330, Brazil}

\author{ Stefano de Gironcoli} 
\affiliation{Scuola Internazionale Superiore di Studi Avanzati (SISSA),
Via Bonomea 265, I-34136 Trieste, Italy}
\affiliation{CNR-IOM DEMOCRITOS Simulation Center, Via Bonomea 265, I-34136 Trieste, Italy}

\begin{abstract}
The Gauge-Including Projector Augmented Wave (GIPAW) method, within
the Density Functional Theory (DFT) Generalized Gradient Approximation
(GGA) framework, is applied to compute solid state NMR parameters for
$^{27}Al$ in the $\alpha$, $\theta$, and $\kappa$ aluminium oxide phases 
and their gibbsite and boehmite precursors. The
results for well-established crystalline phases compare very well with
available experimental data and provide confidence in the accuracy of
the method. For $\gamma$-alumina, four structural models proposed in
the literature are discussed in terms of their ability to reproduce
the experimental spectra also reported in the literature.  Among the
considered models, the $Fd\overline{3}m$ structure proposed by 
Paglia~{\it et~al.} [Phys. Rev. B {\bf 71}, 224115 (2005)] shows the
best agreement.  We attempt to link the theoretical NMR parameters to
the local geometry.  Chemical shifts depend on coordination number but
no further correlation is found with geometrical parameters. Instead our
calculations reveal that, within a given coordination number, a linear
correlation exists between chemical shifts and Born effective charges.

\end{abstract}

\maketitle

\section{Introduction}
\label{Intro}

Aluminium oxide ($Al_2O_3$), also known as alumina, is one
of the most important oxides because of its many industrial
applications\cite{CaiRashkeev2003}.  Corundum ($\alpha$-$Al_2O_3$),
the most stable and common crystalline form of alumina, is the
final product of the calcination of hydroxides or oxyhydroxides of
aluminium at temperatures above 1273 K, which can vary according to
the precursor. The transformation from aluminium hydrates to the final
oxide is not direct and a variety of relatively stable intermediate
phases can be detected for each combination of starting compound
and thermal treatment. 

The most commonly used precursors in the synthesis of the various alumina
phases are gibbsite [$\gamma$-$Al(OH)_3$], bayerite [$\alpha$-$Al(OH)_3$],
and boehmite [$\gamma$-$AlO(OH)$]. 
The transformation of gibbsite to corundum can proceed through a sequence
of hexagonal close packed aluminas ($\chi$ and $\kappa$) or an alternative
sequence in which boehmite is initially formed and the $\alpha$ phase
is achieved via the cubic spinel transition phases ($\gamma$, $\delta$,
and $\theta$)\cite{MacKenzieTemuujin1999}. Bayerite can also follow this
same sequence through boehmite to form corundum\cite{CesterosSalagre1999}
or an alternative path transforming to $\eta$ and $\theta$
phases\cite{Brown_JCS_1953}.

Among the transition aluminas, the $\gamma$ phase is highly valued
for industrial applications due to its textural properties (surface
area, pore volume, pore size), which makes it an important material
in many industrial processes acting as an adsorbent, a catalyst
and/or catalyst support. In petroleum and petrochemical industries
$\gamma$-alumina is used as catalyst support for transition-metal
sulfides Co(Ni)MoS in hydrotreatment catalysts and metallic alloys in
reforming catalysts\cite{Ertl_Book_1997,DigneSautet2004}. Due to its low crystallinity
and the consequent difficulty in characterization, the debate on the
structure of $\gamma$-alumina remains open and a series of theoretical and
experimental works concerning this subject have been published along the
decades\cite{StumpfRussell1950,ZhouSnyder1991,LeeCheng1997,Wang_JPC_1999,Wolverton_PRB_2000,Sohlberg_CEC_2000,KrokidisRaybaud2001,RaybaudDigne2001,Smrcok_AC_2006,FerreiraMartins2011}.

Since the transition between the distinct intermediate phases is a gradual
process, the precise temperature at which each phase is obtained with
a high degree of purity can not be determined by X-ray diffraction (XRD) experiments only.
Solid-State Nuclear Magnetic Resonance (SS-NMR) is an important technique
for material characterization. Long-range order is not a prerequisite
to distinguish different phases and the knowledge of $^{27}Al$ NMR data
can allow the detection of the onset of phase changes during alumina
calcination\cite{HillBastow2007}, permitting a discussion of the
transition mechanisms\cite{ODellSavin2007}. A detailed interpretation
of the results remains, however, a challenge.

The advent of theoretical techniques such as the Gauge-Including Projector
Augmented Wave (GIPAW) method\cite{PickardMauri2001,YatesMauri2007}
enables the {\it ab initio} calculation of isotropic chemical shielding,
$\sigma_{iso}$, quadrupolar coupling constant, $C_Q$, and asymmetry parameter, $\eta_Q$, in solids. 
First principles simulations of NMR spectra of
structural models for transition aluminas are now
possible\cite{ChoiMatsunaga2009,LizarragaHolmstrom2011} and can be
compared with available experimental data.

In this paper we apply the GIPAW method to compute solid state Magic Angle
Spinning (MAS) NMR parameters for a number of well characterized aluminium
oxide phases and for their boehmite and gibbsite precursors. The results
are compared with available experimental data to validate the method. In
order to contribute to a better characterization of $\gamma$-alumina,
a number of structural models, proposed in the literature, are examined
and their simulated spectra compared to experimental ones, thus revealing
their adequacy.

We then examine possible correlations of the predicted chemical shifts
with local atomic geometry or local electronic structure, described
through Bader analysis\cite{Bader} and Born dynamical effective
charges\cite{zstar}, finding significant correlations.

The rest of the paper is organized as follows: in
Sec.~\ref{DetailsCalculations} we describe the theoretical methodology
and the structural models used. In Sec.~\ref{Results} we present our
calculated NMR results, compare them with experiments and discuss the
resulting correlations. Sec.~\ref{conclusions} contains our conclusions.

\section{Details of Calculations}
\label{DetailsCalculations}

\subsection{Electronic structure}
\label{ElectronicStructure}

All {\it ab initio} calculations in this study were performed using the
codes available within the Quantum ESPRESSO distribution\cite{QEspresso},
which implements the DFT\cite{HohenbergKohn1964} framework using a plane
waves basis set to expand the one-electron wavefunctions of Kohn-Sham
equations\cite{KohnSham1965}. The effect of exchange-correlation (XC)
potential was explored by comparing the results of different
descriptions for this term: Perdew-Burke-Ernzerhof (PBE) generalized
gradient approximation\cite{PerdewBurke1996} and its revision
(revPBE) by Zhang and Yang\cite{ZhangYang1998}. Furthermore,
we also considered a van der Waals-aware density functional
(vdW-DF)\cite{DionRydberg2004,ThonhauserCooper2007} recently implemented
in Quantum ESPRESSO. Interaction of valence electrons with nuclei
and core electrons were treated by the Projector Augmented-Wave
(PAW)\cite{Blochl1994} method.

The plane wave kinetic energy cut-off and k-points sampling were
adjusted to yield less than 1 mRy/atom convergence in total energy
for all models. A kinetic energy cut-off of 45 Ry and expansion
of augmentation charges up to 220 Ry was sufficient to ensure this
criterion. Integration in the Brillouin zone were determined by
the Monkhorst-Pack\cite{MonkhorstPack1976} procedure\cite{kmesh}.
Both atomic positions and cell vectors were fully optimized.
NMR chemical shieldings were converged within less than 1 ppm. 

The Born effective charge tensor $Z^*_{\kappa, \alpha\beta}$ is
defined\cite{zstar} by the macroscopic polarization induced, in direction
$\beta$ and under conditions of zero macroscopic electric field, by a
zone center phonon displacing atomic sublattice $\kappa$ in direction
$\alpha$. Effective charges were calculated within density functional
perturbation theory (DFPT)\cite{GiannozziGironcoli1991,DFPT-RMP-2001}
and the values presented in the following sections correspond to
their isotropic component obtained by $Z^*_{\kappa}=Tr[Z^*_{\kappa,
\alpha\beta}/3]$.

\subsection{NMR chemical shifts and quadrupolar parameters}
\label{NMRParameters}

First principles GIPAW calculations\cite{PickardMauri2001,YatesMauri2007}
yield the absolute chemical shielding tensors for each
nucleus, $\vec{\sigma}(r)$. Isotropic chemical shieldings,
$\sigma_{iso}=Tr[\vec{\sigma}/3]$, are compared to the experimental
isotropic chemical shifts by using the standard expression:
$\delta_{iso}=\sigma_{ref}-\sigma_{iso}$. In this work we choose corundum
as reference such that $^{27}Al$ shift of the $\alpha$ phase is aligned
to the experimental one at 0 ppm.

The resulting {\it ab initio} NMR spectra were obtained by using
the QuadFit program\cite{KempaSmith2009} with the theoretically
calculated chemical shifts and quadrupolar interaction parameters,
using the experimental magnetic field intensity, and normalizing each
spectral component to reflect the relative number of aluminium types. A
Lorentzian broadening was added to each spectral feature to obtain the
best comparison with the experimental lineshape.

\subsection{Structural models}
\label{StructuralModels}

Among the oxide phases with well characterized structures, we studied the
final product $\alpha$ and the transitional aluminas $\theta$ and $\kappa$.

The structural model for $\alpha$-alumina, used as a reference for all 
simulated NMR spectra in this work, was published by 
Ishizawa~{\it et~al.}\cite{IshizawaMiyata1980} in an XRD study. The structure was
reported as a corundum-type, with an hexagonal crystal system and
$R\overline{3}c$ space group. The crystallographic cell contains six $Al_2O_3$
units in which all aluminium sites are coordinated by six oxygens. 

The $\theta$ phase is present along the $\gamma \rightarrow
\alpha$ transition in different dehydration paths and its 
structure has been characterized by Zhou and Snyder\cite{ZhouSnyder1991},
with a monoclinic crystal system and $C2/m$ space group. In that work 
the structure was refined from Rietveld analysis resulting in a 
crystallographic cell with four $Al_2O_3$ units, in which half of the
aluminium atoms are octahedrally ($Al_{oct}$) and half are
tetraedrally ($Al_{tet}$) coordinated.

The $\kappa$ phase is one of the intermediate products of the
dehydration path from gibbsite to corundum.  Ollivier~{\it et~
al.}\cite{OllivierRetoux1997} describes the $\kappa$ phase with an
orthorhombic system and $Pna2_1$ space group. The cell contains six
$Al_2O_3$ units and 25\% of the $Al^{3+}$ sites are tetrahedral, 50\%
octahedral and 25\% in a very distorted octahedral.

We have also studied the NMR spectra of precursor phases gibbsite and boehmite. 
For gibbsite we started from the structure resolved by Saalfeld and Wedde
in a single-crystal XRD study\cite{SaalfeldWedde1974}. The structure
is monoclinic with $P2_1/n$ space group where the $Al^{3+}$ cations are
octahedrally coordinated by 6 $OH^{-}$ groups forming double layers
and occupy two thirds of the octahedral holes in alternate layers\cite{GallardoZK1996}. 
The interlayer cohesion is granted by
hydrogen bonds between these $OH^{-}$ groups.

In the also layered boehmite structure, described by 
Christensen~{\it et~al.}\cite{ChristensenLehmann1982}, 
each $Al^{3+}$ cation
is octahedrally coordinated by 2 $OH^{-}$ groups and 4 intralayer
$O^{2-}$ anions in an orthorhombic system with $Cmcm$ space group.
As in the gibbsite structure, the boehmite double layers interact with
each other via hydrogen bonds, which are exclusively interlayer. In
this phase, hydrogen bonds are organized in chains along [001]
direction and different bond networks are possible depending on the
relative orientation of neighboring chains. In our calculations we
considered all possible combinations compatible with a 2$\times$2
supercell. The structure with the lowest energy was found to be the
one where nearest neighbor chains are antiparallel. However energy
differences among various combinations are found to be less than 1
mRy/cell which is consistent with the experimentally observed disorder
in this phase\cite{DamodaranRajamohanan2002}.

As one of the aims of this work is to contribute to the characterization
of $\gamma$-alumina phase, four different $\gamma$-alumina structural
models from published theoretical works have been studied.

The first $\gamma$-phase model, $\gamma$-$Al_2O_3(A)$,
considered here was proposed by 
Gutierrez~{\it et~al.}\cite{GutierrezTaga2001,Menendez-ProupinGutierrez2005} This model,
also called {\it defect spinel} or {\it spinel-like} structure,
consists of a cell with 8 $Al_2O_3$ units, in which 37.5\% of the cations
are $Al_{tet}$ and 62.5\% are $Al_{oct}$. Among the 24 O atoms, 12 are
four-fold coordinated ($O_{4-fold}$) and 12 are three-fold coordinated
($O_{3-fold}$) to aluminium atoms\cite{ChingOuyang2008}.
It is important to note that in this model only spinel sites are 
occupied by the $Al^{3+}$ cations.

The next $\gamma$-phase model, $\gamma$-$Al_2O_3(B)$, used in
this work was published by Digne~{\it et~al.}\cite{DigneSautet2004}
and proposed by Krokidis~{\it et~al.}\cite{KrokidisRaybaud2001}. 
The model has 8 $Al_2O_3$ units in the cell,
25\% of all aluminium atoms are $Al_{tet}$ sites, in a sublattice of
$O^{2-}$ anions. The crystal system is monoclinic, but very close to an
orthorhombic one with the $P2_1/m$ space group. 
In this model cations occupy also non-spinel sites.

Two other $\gamma$-phase models were published by 
Paglia~{\it et~al.}\cite{Paglia2004,PagliaRohl2005}. 
The unit cells of these models contain
a large number of atoms, 32 $Al_2O_3$ units, and were
generated from an extensive search on all structural possibilities of
the $\gamma$-$Al_2O_3$ structures using $Fd\overline{3}m$ and $I4_1/amd$
space groups. In the $Fd\overline{3}m$ model, $\gamma$-$Al_2O_3(C)$,
among the 64 $Al^{3+}$ sites 22 are $Al_{tet}$, 41 are $Al_{oct}$ and
1 is $Al_{pen}$ (five-coordinated), while in the $I4_1/amd$ model,
$\gamma$-$Al_2O_3(D)$, there are 21 $Al_{tet}$ and 43 $Al_{oct}$
sites. In these two models, due to the breaking of the local symmetry
by the variations in cation occupancies and related distortions
in octahedral and tetrahedral sites, the symmetry is actually
$P_1$\cite{LoyolaMenendez-Proupin2009}.

\section{Results and discussion}
\label{Results}
\subsection{Structure optimization and approximations on exhange-correlation
functional}
\label{StructureOpt}

Geometry optimization was performed for all phases, allowing both the
atomic positions and cell vectors to relax keeping the group symmetry
fixed. To investigate the effect of approximations on exhange-correlation
functionals, each geometry optimization was repeated with PBE, rev-PBE,
and the vdW-DF functionals as described in Sec.~\ref{ElectronicStructure}.
All XC functionals considered overestimate the experimental volume,
PBE by about 1.8\% on average, revPBE by 4.2\%. The use of van der Waals
functional was found to have negligible effect, inducing a tiny further
expansion, on all structures except for gibbsite ($\gamma$-$Al(OH)_3$)
[see Table~\ref{tbl:volumes}] where a slight contraction was observed. The
exception in the case of gibbsite can be understood considering its
layered and open structure.

\begin{table}[htb]
 \centering
 \caption{Volumes of the experimental and optimized cells with PBE,
rev-PBE, and vdW-DF functionals for the five phases with well characterized
experimental structures considered in this work. }
 \begin{tabular}{ccccc}
 \hline
 & \multicolumn{4}{c}{Volume (\AA{}$^3$)}\\
 Structure & Experiment & PBE & rev-PBE & vdW-DF\\
 \hline
 $\alpha$-$Al_2O_3$ & 254.25$^a$ & 261.84 & 266.30 & 268.06\\
 $\theta$-$Al_2O_3$ & 187.92 $^b$& 192.53 & 195.54 & 196.34\\
 $\kappa$-$Al_2O_3$ & 361.31$^c$ & 369.19 & 375.21 & 377.75\\
 $\gamma$-$AlO(OH)$& 261.13$^d$ & 262.61 & 270.24 & 271.65\\
 $\gamma$-$Al(OH)_3$ & 427.98$^e$ & 431.57 & 449.51 & 447.30\\
 \hline 
 \multicolumn{5}{l}{\begin{small}
$^a$Ref.~\on-line-cite{IshizawaMiyata1980} $^b$Ref.~\on-line-cite{ZhouSnyder1991}.
$^c$Ref.~\on-line-cite{OllivierRetoux1997}. $^d$Ref.~\on-line-cite{ChristensenLehmann1982}.
$^e$Ref.~\on-line-cite{SaalfeldWedde1974}.\end{small}}\\
 \end{tabular}
 \label{tbl:volumes}
\end{table}

Since {\it ab initio} NMR calculations are very sensitive to structural
details, we performed all NMR calculations at optimized positions for
each XC functional. In spite of the discrepancy in calculated equilibrium
volume, the average difference in $\sigma_{iso}$, $|C_Q|$ and $\eta_Q$ 
between PBE and revPBE were calculated as 1.15 ppm, 0.14 MHz and 0.03, 
respectively. 
These values were found to have negligible effect on total 
spectra and, except for gibbsite, only the PBE spectra will be shown.
The numerical values for the three XC functionals are reported in Table~\ref{tbl:nmr-crystals}.

\subsection{NMR parameters and spectra for well characterized structures}
\label{NMRWellDescribedStructures}

In this section, we present a detailed comparison of our {\it ab
initio} calculated chemical shifts and quadrupolar interaction
parameters for all the well characterized structural phases mentioned
in Sec.~\ref{StructuralModels} with experimental data, as well as with
very recent theoretical results from the literature. All relevant data
are shown in Table~\ref{tbl:nmr-crystals}.


\begin{small}
\begin{table*}[!ht]
 \caption{Comparison between
chemical shifts and quadrupolar coupling parameters calculated in this
work and experimental and theoretical data from the literature. }

 \begin{tabular}{l ccc ccc ccc}
 \hline
  &  
  \multicolumn{3}{c}{\underline{\hspace{2truecm} $\delta_{iso}$ (ppm) \hspace{2truecm}}} & 
  \multicolumn{3}{c}{\underline{\hspace{2truecm} $C_Q$ (MHz) \hspace{2truecm}}} & 
  \multicolumn{3}{c}{\underline{\hspace{2truecm} $\eta_Q$ \hspace{2truecm}}} \\
  Structure & 
   \multicolumn{2}{c}{previous} & this work & 
   \multicolumn{2}{c}{previous} & this work & 
   \multicolumn{2}{c}{previous} & this work \\
  ~~~Al site   & 
   Exp. & Th. & \begin{tiny} PBE/rev-PBE/vdW-DF \end{tiny} &
   Exp. & Th. & \begin{tiny} PBE/rev-PBE/vdW-DF \end{tiny} &
   Exp. & Th. & \begin{tiny} PBE/rev-PBE/vdW-DF \end{tiny} \\
 \hline
  $\alpha$-$Al_2O_3$ \\
 ~~~$Al_{oct}$ &
   0.0 (1)$^a$ & 0.0$^b$ & 0.0 & 
   2.38$^a$ & 2.33$^b$& 2.05/1.98/2.10 & 
   0.00$^a$ & 0$^b$& 0.04/0.04/0.03  \\
   \\
 $\theta$-$Al_2O_3$ \\
 ~~~$Al_{oct}$ & 
 -3.0 (1.0)$^a$ & -5.9$^b$ & -4.9/-4.8/-5.0& 
  3.50 (0.30)$^a$ & 3.44$^b$ & 2.91/2.81/2.71 & 
  0.00 (0.10)$^a$ & 0.18$^b$ & 0.26/0.36/0.29 \\

 ~~~$Al_{tet}$ & 
  66.5 (1.0)$^a$ & 62.3$^b$ & 62.6/60.2/61.1 & 
  6.40 (0.10)$^a$ & 6.34$^b$ & 5.63/5.53/5.37 & 
  0.65 (0.02)$^a$ & 0.42$^b$ & 0.43/0.37/0.42 \\
   \\
  $\kappa$-$Al_2O_3$ \\
 ~~~$Al_{oct} (1)^f$ & 
   0.0$^c$ &  0.3$^{b}$ & 2.0/2.3/2.0 & 
  0.50$^c$ &  -9.98$^{b}$  & -9.53/-9.45/-9.41 &
         -- &  0.33$^{b}$ & 0.29/0.29/0.28 \\
 ~~~$Al_{V + I} (2)^f$ & 
          --  & 1.4$^{b}$  & 7.1/6.7/6.8 & 
 $>$ 1.5$^c$ & 5.20$^{b}$ & 4.43/4.33/4.27 & 
           -- & 0.94$^{b}$  & 0.76/0.77/0.68 \\
 ~~~$Al_{tet} (3)^f$ &
  68.5$^c$ & 60.7$^{b}$ & 62.6/60.2/61.2 &
  0.76$^c$ & -5.53$^{b}$ & -4.83/-4.67/-4.49 &
   0.3$^c$ & 0.33$^{b}$ & 0.33/0.31/0.32 \\
 ~~~$Al_{oct} (4)^f$ &
   5.0$^c$ & 4.4$^{b}$ & 3.8/3.6/3.5 & 
  0.85$^c$ & 4.51$^{b}$ & 4.99/-4.96/5.06 & 
         -- & 0.77$^{b}$ & 0.99/1.00/1.00 \\
   \\
  \multicolumn{2}{l}{ $\gamma$-$Al(OH)_3$ Gibbsite} \\
 ~~~$Al_{oct}(1)$ & 
   -5.6$^{d,g}$ & 2.2$^e$ & 2.2/2.3/2.1$^h$ &
   4.70$\pm$0.20$^d$ & 5.10$^e$& -5.30/-4.74/-4.43 & 
   1.00$\pm$0.05$^d$ & 0.35$^e$ & 0.32/0.35/0.35 \\ 
 ~~~$Al_{oct}(2)$ & 
   0.0$^{d,g}$ & 0.0$^e$ & 0.0$^h$ &
   2.20$\pm$0.20$^d$ & 2.80$^e$& 2.33/2.06/1.81 & 
   0.75$\pm$0.05$^d$ & 0.66$^e$ & 0.81/0.71/0.66 \\ 
   \\
 \multicolumn{2}{l}{$\gamma$-$AlO(OH)$ Boehmite} \\
 ~~~$Al_{oct}$ &
  -1.0$^{d,g}$ & -- & -0.9/-0.5/-0.2$^h$ & 
  1.8 - 2.8$^d$ & -- & 2.16/2.08/2.27 &
  0.5 - 1.0$^d$ & -- & 0.45/0.46/0.67 \\ 
 \hline
\multicolumn{10}{l}{\begin{small}
$^a$Ref.~\on-line-cite{ODellSavin2007}; 
$^b$Ref.~\on-line-cite{LizarragaHolmstrom2011};
$^c$Ref.~\on-line-cite{OllivierRetoux1997}; 
$^d$Ref.~\on-line-cite{DamodaranRajamohanan2002};
$^e$Ref.~\on-line-cite{VyalikhZesewitz2010}.\end{small}}\\
\multicolumn{10}{l}{\begin{small}
$^f$ Aluminium sites are labeled according to Ref.~\on-line-cite{OllivierRetoux1997} and the sites symmetry, see Table~\ref{tbl:kappa-distortion}.\end{small}}\\
\multicolumn{10}{l}{\begin{small}
$^g$ The experimental $\delta_{iso}(Al_{oct}(2))$ = 11.5$\pm$0.2 ppm from gibbsite is taken as reference.\end{small}}\\
\multicolumn{10}{l}{\begin{small}
$^h$ The calculated $\sigma_{iso}(Al_{oct}(2))$ = 562.72 ppm from gibbsite is taken as reference.\end{small}}\\
 \end{tabular}
 \label{tbl:nmr-crystals}
\end{table*}
\end{small}


\subsubsection{$\alpha$ and $\theta$ phases}

Corundum, $\alpha$-$Al_2O_3$, shows a single well defined NMR peak that we take
as reference when comparing spectra for other structures.

\begin{figure}[!]
 \centering
 \includegraphics[width=\columnwidth]{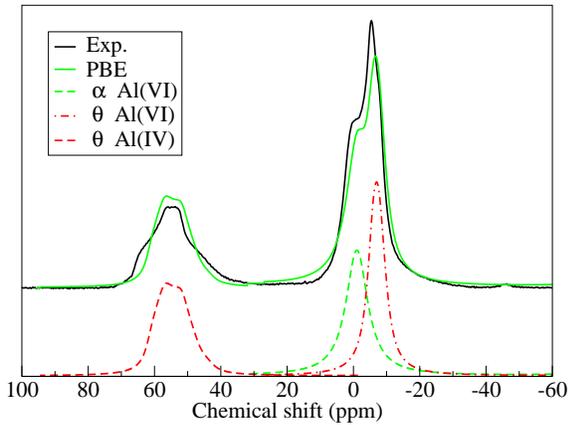}
 \caption{(color on-line) Comparison between our theoretical results
and the experimental $^{27}Al$ MAS NMR spectra from
Ref.~\protect\on-line-cite{ODellSavin2007} for a mixture of the $\alpha$ and
$\theta$ alumina phases (upper curves). Decomposition of the theoretical
spectrum in its individual components (lower curves).}

 \label{fig:nmr-corundum-theta}
\end{figure}

For the $\theta$ phase, we compare our results with the ones reported
in Ref.~\on-line-cite{ODellSavin2007}, where a mixture of $\alpha$
and $\theta$ phases were examined, and the NMR parameters (see
Table~\ref{tbl:nmr-crystals}) for these two coexisting phases were
extracted from the experimental spectrum by lineshape fitting.

The overall good agreement between experimental and theoretical spectra
can be seen in Fig.~\ref{fig:nmr-corundum-theta}, upper curves.
The decomposition in individual contributions is also given.
The main difference between experiment and theory is in the asymmetry
parameter $\eta_Q$ for the $Al_{tet}$ site. For this peak, it should
be noted that 
a recent theoretical work\cite{LizarragaHolmstrom2011}
has also reported 
a value similar to ours.


\subsubsection{$\kappa$ phase}

\begin{table}[!]
 \centering
 \caption{Optimized $Al-O$ distances for the $\kappa$-$Al_2O_3$ structure.
Additionally, for the experimental\protect\cite{OllivierRetoux1997} and the
optimized structures, we present the average absolute deviation in the
distances, $D_{dist.}$, and in the $O-Al-O$ angles, $D_{ang.}$, according
to equations~(\ref{eq:dev-distances}) and~(\ref{eq:dev-angles}).}
 \begin{tabular}{ccccc}
 \hline
  & $Al_{oct} (1)$ & $Al_{V + I} (2)$ & $Al_{tet} (3)$ & $Al_{oct} (4)$ \\
 \hline
 Distances (\AA{}) & 1.962 & 1.991 & 1.797 & 1.879 \\
  & 1.961 & 2.269 & 1.772 & 1.959 \\
  & 1.936 & 1.848 & 1.756 & 1.821 \\
  & 1.820 & 1.829 & 1.784 & 2.026 \\
  & 1.917 & 2.038 & -- & 1.838 \\
  & 1.968 & 1.844 & -- & 2.215 \\\\
 $D_{dist.}$ (\AA{}) & & & & \\
 Th.  & 0.039 & 0.130 & 0.013 & 0.111 \\
 Exp. & 0.058 & 0.171 & 0.025 & 0.105 \\\\
 $D_{ang.}$ (\textdegree) & & & & \\
 Th.  & 5.028 & 8.392 & 3.572 & 7.305 \\
 Exp. & 5.150 & 8.590 & 5.081 & 6.888 \\
 \hline
 \end{tabular}
 \label{tbl:kappa-distortion}
\end{table}

\begin{figure}[!]
 \centering
 \includegraphics[width=\columnwidth]{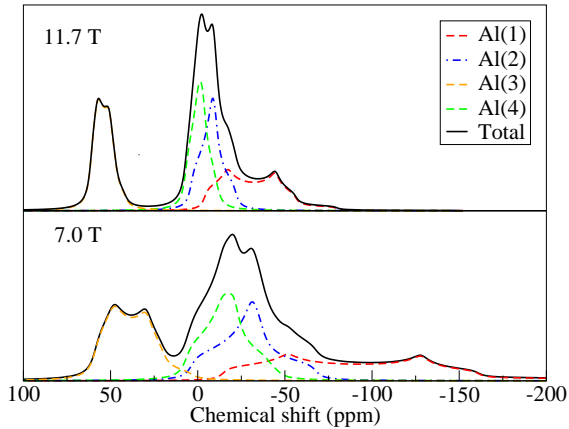}
 \caption{(color on-line) Simulated $^{27}$Al MAS NMR spectra for
$\kappa$-alumina at 11.7 T (upper graph) and 7.0 T (lower graph).
Decomposition of the spectra in individual components is also given.}
 \label{fig:nmr-kappa}
\end{figure}
 
The crystal structure of $\kappa$-alumina has been reported in a work
by Ollivier~{\it et~al.}\cite{OllivierRetoux1997}, 
in which
$^{27}Al$ MAS NMR and multiple quantum magic angle spinning (3Q MQ MAS) 
$^{27}Al$ NMR experiments were also performed and interpreted 
on the basis of the local structure.
Three distinct resonance signals were indentified in the spectrum obtained 
with magnetic field intensity of 
7.0~T, two of them from $Al_{oct}$ sites. At higher magnetic field of 11.7~T, 
further resolution revealed an additional peak for this site at
high shielding, resulting in the interpretation of 
the complex lineshape for $Al_{oct}$ peak region as the overlap of
three distinct peaks: two with high, one with low quadrupolar coupling.
The high coupling component was assigned to the strongly distorted
$Al_{oct}$ site, $Al (2)$ in the published crystallographic description,
and was labeled as $Al_{V+I}$ to stress the presence of a strongly
elongated $Al-O$ bond.

In order to discuss the local distortion of the aluminium sites we
calculated the average absolute deviation in distances
\begin{equation}
\label{eq:dev-distances}
 D_{dist.} = \sum_{i=1}^{n}\frac{|d_i-\overline{d}|}{n}
\end{equation}
where n = 4 or 6 for $Al_{tet}$ and $Al_{oct}$ 
sites, respectively. The angular deviation of the $O-Al-O$ angles
from their ideal values in octahedra and tetrahedra are calculated as
\begin{equation}
\label{eq:dev-angles}
 D_{ang.} = \sum_{i=1}^{k}\frac{|\beta_{i}-\beta_{ref}|}{k}
\end{equation}
where k = 6 and $\beta_{ref}$ = 109.47\textdegree\ for the $Al_{tet}$ sites
or k = 12 and $\beta_{ref}$ = 90\textdegree\ for the $Al_{oct}$ sites.

As it can be seen in Table~\ref{tbl:kappa-distortion}, our calculations
confirm the order of increasing distortion in octahedral sites suggested
in the experimental analysis and in particular site $Al_{V+I}(2)$
is found to be the most distorted.

The simulated total spectra, reported in Fig.~\ref{fig:nmr-kappa},
obtained with our calculated $\delta_{iso}$, $C_Q$ and $n_Q$, are in
very good qualitative agreement with the $^{27}Al$ MAS NMR experimental
spectrum of Ref.~\on-line-cite{OllivierRetoux1997} at both magnetic
field intensities.

However, from the theoretical decomposition of the composite octahedral
peak, it can be seen that the high shielding feature around -50 ppm, which
was experimentally observed at high magnetic field intensity
and was assigned to be Al(2) on the basis of the distortion data, is
actually part of a bimodal peak with high quadrupolar coupling and low
asymmetry parameter, which belongs to Al(1). Morever, the experimental detection
of this particular feature only at high field intensity can also be
understood considering the {\it ab initio} spectrum for low field intensity,
reported in the lower part of Fig.~\ref{fig:nmr-kappa}. As can be seen,
at low magnetic field intensity, Al(1) peak broadens considerably due to
its high quadrupolar coupling constant, making it diffcicult to observe
over the background in the experiment.
This demonstrates that, even though experiments can provide accurate
structural information, {\it ab initio} NMR calculations might be
essential for an unambiguous peak assignment.


\subsubsection{gibbsite and boehmite phases}
 \label{sec:boehmite-and-gibbsite}

The structures and NMR properties of
gibbsite and boehmite are experimentally well
characterized.\cite{SladeSouthern1991-a,DamodaranRajamohanan2002,AshbrookMcManus2000,VyalikhZesewitz2010}.

\begin{figure}[htb]
 \includegraphics[width=\columnwidth]{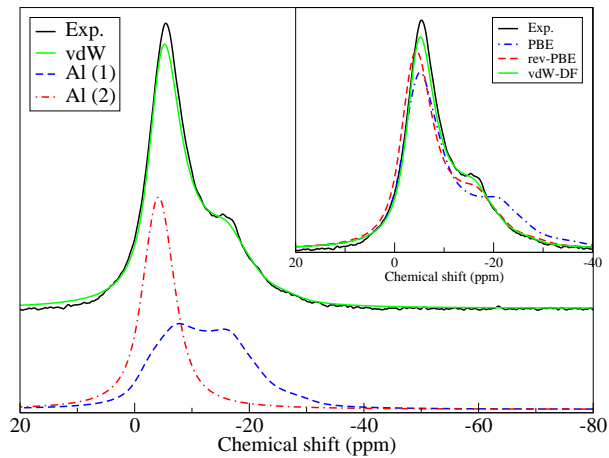}
 \caption{(color on-line) Comparison between the experimental $^{27}$Al
 MAS NMR spectra from Ref.~\protect\on-line-cite{HillBastow2007} for gibbsite and the
 theoretical spectrum obtained with the vdW-DF functional.  Decomposition
 of the spectrum in individual components is also given.  The effect of
 different exchange and correlation functionals on the simulated spectrum
 is shown in the inset. }

 \label{fig:nmr-gibbsite}
\end{figure}

As can be seen in Fig.~\ref{fig:nmr-gibbsite}, the gibbsite NMR
spectrum simulated with the theoretically determined parameters agrees
very satisfactorily with the experimentally obtained spectrum by 
Hill~{\it et~al.}\cite{HillBastow2007}.  This structure is the only
one for which we observed a noticeable dependence of the theoretical
spectra on the exchange correlation functional used, showing  improved
results when using vdW-DF\cite{DionRydberg2004} functional (see inset in
Fig.~\ref{fig:nmr-gibbsite}). In Table~\ref{tbl:nmr-crystals},
we present a quantitative comparison between our calculated NMR
parameters and the experimental results reported by Damodaran {\it et
al.}~\cite{DamodaranRajamohanan2002}. Our results for $C_Q$ and $\eta_Q$
are in good agreement with both of these experiments and can accurately
reproduce the asymmetry of the peak.

In agreement with the previous theoretical work by Vyalikh {\it et
al.}~\cite{VyalikhZesewitz2010}, in Fig.~\ref{fig:nmr-gibbsite} we show
that this resonance profile can be decomposed as the superposition of
two distinct peaks with different second-order quadrupolar structure,
belonging to two distinct $Al_{oct}$ sites in gibbsite.

\begin{figure}[htb]
 \centering
 \includegraphics[width=\columnwidth]{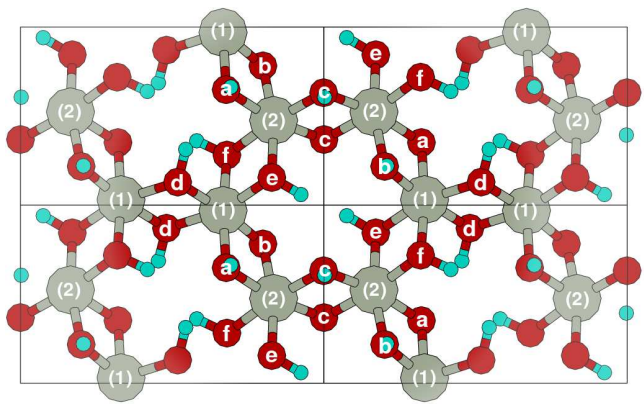}
 \caption{(color on-line) View from direction (001) of a single Al(OH)$_3$
 layer in gibbsite showing the arrangement of the different types of
 hydroxyl groups around the aluminium cations (large gray atoms). Oxygen
 (small red atoms) are labeled according to their different environments
 as referred in Table~\ref{tbl:oxygens-gibbsite}.}

 \label{fig:aluminium-octahedra}
\end{figure}

As described in Ref.~\on-line-cite{VyalikhZesewitz2010} the two
sites differ in the OH groups surrounding them. This can be seen
in Fig.~\ref{fig:aluminium-octahedra} where a single
Al(OH)$_3$ layer is drawn showing that among the six OH groups 
surrounding $Al_{oct}(1)$, two participate in interlayer hydrogen-bonds as donors 
while the remaining four are oriented in-plane and participate to
interlayer hydrogen-bonds as acceptors. For $Al_{oct}(2)$ the opposite occours.

We further characterize the two aluminium sites reporting, in
Table~\ref{tbl:oxygens-gibbsite}, the $^{17}O$ chemical shifts, the Born
effective charges and the hydrogen-bond connectivity of the six oxygen
types (labeled in Fig.~\ref{fig:aluminium-octahedra}) surrounding them.
This analysis reveals that the two aluminium sites are mostly surrounded
by the same types of oxygens and the distinction is based on just two
different oxygen environments ($O_{c}$ and $O_{d}$), each one neighboring
only one type of aluminium site.

\begin{table}[htb]
 \centering
 \caption{ Effective charges, chemichal shifts and hydrogen-bonding
 properties (ID = interlayer donor; IA = interlayer acceptor) for the different
 types of oxygen atoms in gibbsite.  Labelling according to
 Fig.~\ref{fig:aluminium-octahedra}.}
 \begin{tabular}{cccc}
 \hline
 Site & $Z^*$ & $\sigma_{iso}$ & H-Bond\\
 \hline
 O$_{a}$ &~ -1.46~ &~ 238.2 ~& ID\\
 O$_{b}$ &~ -1.45~ &~ 242.4 ~& ID\\
 O$_{c}$ &~ -1.43~ &~ 246.2 ~& ID\\
 O$_{d}$ &~ -1.41~ &~ 229.4 ~& IA\\
 O$_{e}$ &~ -1.38~ &~ 238.4 ~& IA\\
 O$_{f}$ &~ -1.36~ &~ 228.0 ~& IA\\
 \hline
 \end{tabular}
 \label{tbl:oxygens-gibbsite}
\end{table}

The oxyhydroxide polymorph boehmite, $\gamma$-$AlO(OH)$, was detected
by XRD in the experiment by Hill~{\it et~al.}\cite{HillBastow2007}
in the temperature range from 200 to 400\textdegree C in a mixture
with gibbsite or the transitional $\chi$ phase. A direct comparison
between a simulated NMR spectrum for this phase with that experiment
is therefore not possible. In the experiment by Damodaran {\it et
al.}~\cite{DamodaranRajamohanan2002} the lack of high resolution in the
boehmite NMR spectrum is suggested to be due to the disorder in Al positions in the
sample used, leading to small variations in the isotropic chemical shift
and quadrupolar couplings. Nevertheless, it was possible to extract experimental
estimates for $C_Q$ and $\eta_Q$, which are in good agreement with our
calculated values, as presented in Table~\ref{tbl:nmr-crystals}.


\subsection{The $\gamma$-$Al_2O_3$ phase and its structural models}
\label{GAluminaNMR}

\begin{figure*}[!]
 \includegraphics[width=\columnwidth]{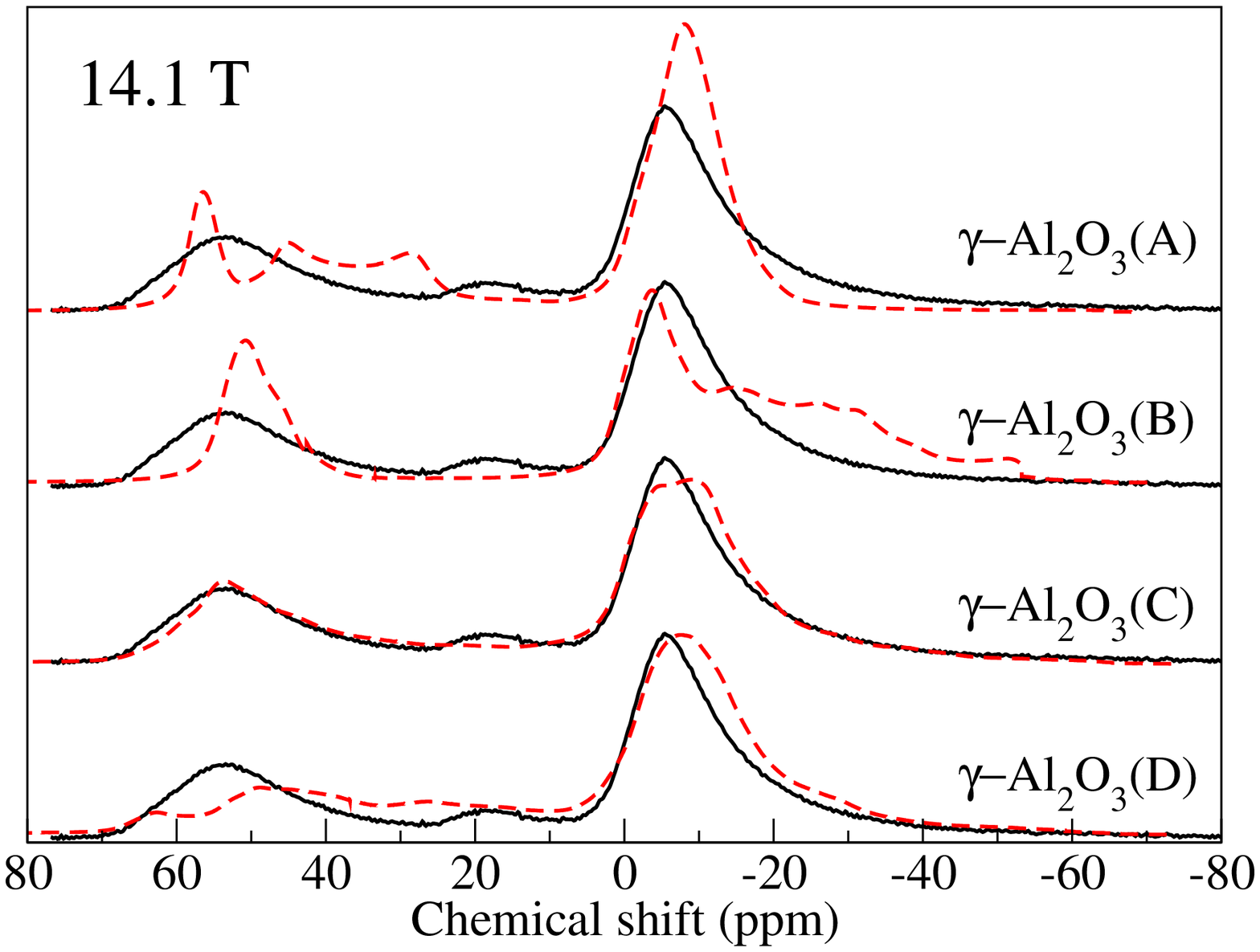}
 \includegraphics[width=\columnwidth]{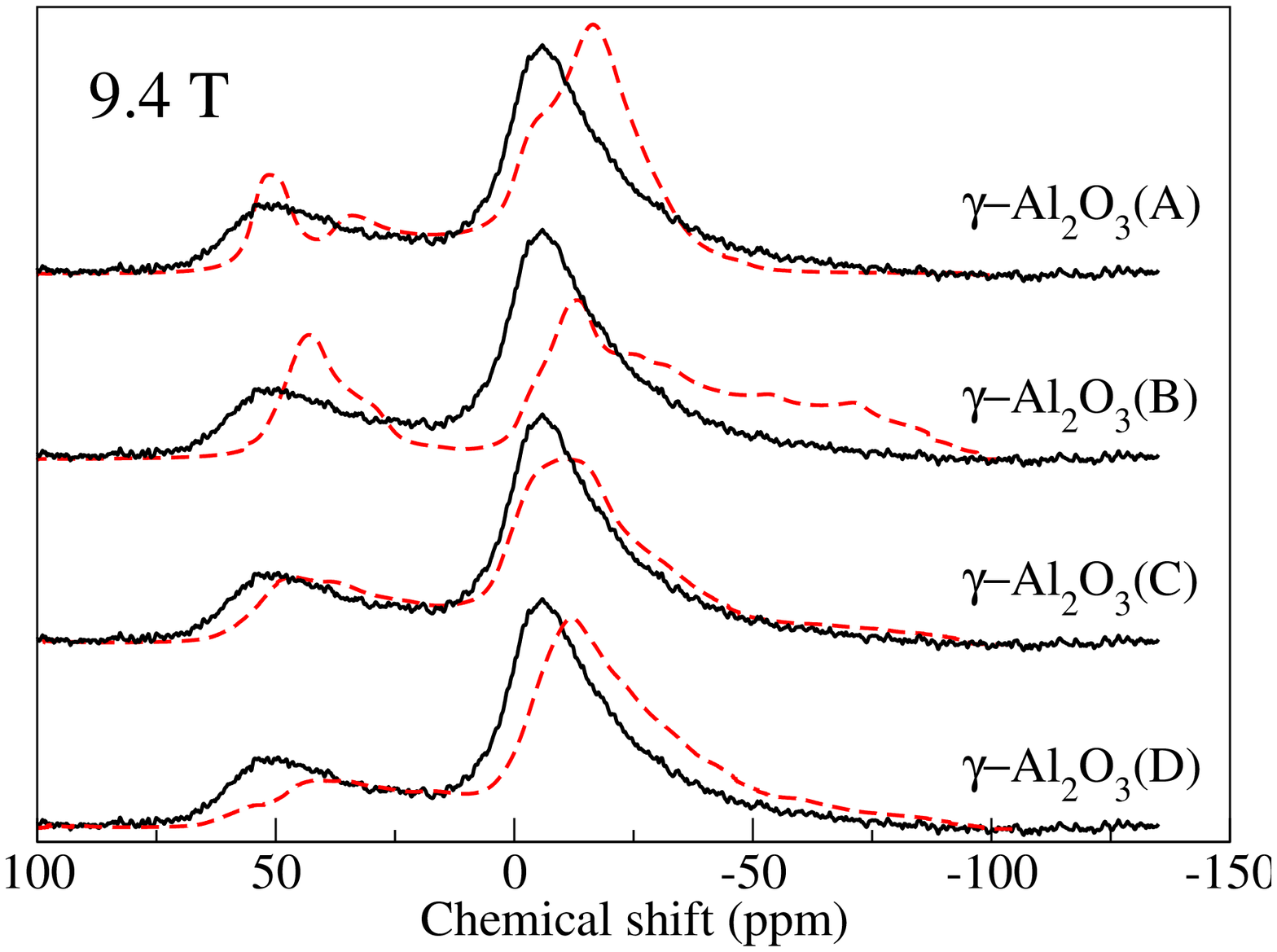}
 \caption{(color on-line) Comparison of the theoretical spectra (dashed
 red lines) for the four $\gamma$-alumina models considered in this
 study and the experimental $^{27}$Al MAS NMR spectra (black solid
 lines) from Ref.~\protect\on-line-cite{ODellSavin2007} (left panel) and
 Ref.~\protect\on-line-cite{HillBastow2007} (right panel). A Lorentzian broadening
 of 0.5 KHz was added to each component which were properly normalized
 to reflect relative site abundance.}

 \label{fig:gamma-aluminas}
\end{figure*}

As previously mentioned, $\gamma$-alumina structure is not yet well
characterized in the literature and a vast discussion about this subject
persists. 
Experimental NMR spectra of $\gamma$-alumina have been reported
in a number of studies.  
In the work by O'Dell~{\it et~al.}\cite{ODellSavin2007}, 
the samples obtained with
calcination temperatures from 600 \textdegree C to 1000 \textdegree
C are attributed to mixtures of cubic spinel transitional phases,
as indicated by the presence of a tetrahedral peak around 60
ppm (taking the $\alpha$ phase as a reference).
The fivefold-coordinated peak that has been observed in this phase has been
attributed to inherent disorder~\cite{ODellSavin2007} or surface atom
contributions~\cite{KwakHu2007}. 
The MAS NMR study of Pecharrom\'{a}n~{\it et~al.}\cite{PecharromanSobrados1999} reported
a small percentage
of $AlO_5$ sites as well, along with an occupation of 76.3\% and 21\%
for $Al_{oct}$ and $Al_{tet}$ sites, respectively.
However, no evidence of a fivefold-coordinated Al peak is present 
in some other experiments, such as the one of Ref.~\on-line-cite{PagliaBuckley2003}, obtained from
a hydrogenated boehmite precursor calcinated at 600 \textdegree C for
several hours and the one of 
Ref.~\on-line-cite{HillBastow2007}, in which the sample was 
obtained from the calcination of gibbsite
at 700 \textdegree C through the formation of boehmite at 300 \textdegree
C.

To gain further insight on the structure of this phase, we have calculated the 
{\it ab initio} NMR
parameters and obtained the simulated spectrum for each one of the four $\gamma$-alumina 
structural models that have been mentioned in Section~\ref{StructuralModels} .
From the comparison of the calculated spectra with the experimental ones 
obtained for calcination conditions associated to $\gamma$-alumina, 
an evaluation of the adequacy of these structural models will be possible.
Since NMR is a sensitive probe of the local structural and chemical environment,
comparisons based on NMR parameters will provide complementary constraints with respect to X-ray and
neutron diffraction methods used, for instance, in the detailed structural
search of Ref.~\on-line-cite{PagliaRohl2005}

In Fig.~\ref{fig:gamma-aluminas} we compare the theoretical spectra with
the experimental data by O'~Dell~{\it et~al.}\cite{ODellSavin2007}
(left panel, 14.1~T), and by Hill~{\it et~al.}\cite{HillBastow2007}
(right panel, 9.4~T). We also compared with the 9.4~T spectrum
by Pecharrom\'{a}n~{\it et~al.}~\cite{PecharromanSobrados1999}
(not shown) which agrees very well with the one by Hill~{\it et~al.}.

From Fig.~\ref{fig:gamma-aluminas}, it is clear that the simulated MAS
NMR spectra for the $\gamma$-$Al_2O_3 (C)$ structural model proposed
by Paglia~{\it et~al.}\cite{PagliaRohl2005} best reproduces all the
experimental results considered. The agreement is very satisfactory and
even better for the more precise spectrum at the higher magnetic field
[left panel of Fig.~\ref{fig:gamma-aluminas} ].

\begin{table}[!htb]
 \centering
 \caption{Average values and corresponding standard deviations for
$^{27}Al$ chemical shifts, absolute quadrupolar coupling, asymmetry
parameter, and site occupation, decomposed according to Al coordination
number, for the four $\gamma$-alumina models considered in this work.}

 \begin{tabular}{ccccr}
 \hline
  & $\delta_{iso}$ (ppm) & $|C_Q|$ (MHz) & $\eta_Q$ & \% \\
 \hline
 $\gamma$-$Al_2O_3 (A)$ & & & &\\
 $AlO_6$ & -0.6$\pm$1.8 & 4.90$\pm$0.47 & 0.86 $\pm$ 0.12 & 62.5\\
 $AlO_4$ & 52.1$\pm$3.0 & 6.36$\pm$2.65 & 0.14 $\pm$ 0.08 & 37.5\\
 $\gamma$-$Al_2O_3 (B)$ & & & &\\
 $AlO_6$ & -0.7$\pm$4.9 & 7.98$\pm$3.25 & 0.41$\pm$0.23 & 75~~\\
 $AlO_4$ & 52.0$\pm$0.1 & 4.51$\pm$0.91 & 0.47$\pm$0.35 & 25~~\\
 $\gamma$-$Al_2O_3 (C)$ & & & &\\
 $AlO_6$ & -0.8$\pm$4.3 & 5.93$\pm$2.45 & 0.52$\pm$0.26 & 64.0 \\
 $AlO_5$  & 15.9 & 7.43 & 0.34 & 1.6\\
 $AlO_4$ & 59.9$\pm$4.7 & 7.70$\pm$3.00 & 0.64$\pm$0.25 & 34.4\\
 $\gamma$-$Al_2O_3 (D)$ & & & & \\
 $AlO_6$ & -0.1$\pm$4.2 & 6.25$\pm$2.38 & 0.57$\pm$0.23& 67.2\\
 $AlO_4$ & 52.7$\pm$6.5 & 8.27$\pm$2.92 & 0.58$\pm$0.27 & 32.8\\
 \hline 
 \end{tabular}
 \label{tbl:nmr-paglia}
\end{table}

It may appear natural that the two models that display lesser agreement
with the experimental spectra are the two (models A and B) whose unit
cells only contain 8 $Al_2O_3$ formula units and a reduced number
of non-equivalent aluminium environments, while model C contains a
large number of $Al_2O_3$ formula units allowing a distribution of
NMR parameters, which is more suitable for the representation of the
broad features observed experimentally. Notice however that having a
large number of non-equivalent aluminium environments may be considered
a necessary, but not a sufficient condition to properly reproduce
the experimental NMR spectra. In fact both Paglia's structural
models (C and D) contain 64 $Al_2O_3$ formula units and reproduce neutron
diffraction data~\cite{PagliaRohl2005} equally well but only model
C (the one associated to $Fd\overline{3}m$ symmetry) satisfactorily
reproduces the spectral region of the tetrahedrally coordinated Al atoms
and even for this model the sharpness of the octahedral-Al peak is not
completely satisfactory. 
Table~\ref{tbl:nmr-paglia} shows that, on average, model C gives a
higher average value for $\delta_{iso}$ for the $Al_{tet}$ peak while
the other models display rather similar values that underestimate
the experiment.

Furthermore, in the $\gamma$-$Al_2O_3$ structural model C, a truly
$AlO_5$ site is evident, and its calculated $\delta_{iso}$ of 15.9 ppm
is easily separated from the peaks of octahedrally (-9.4 to 9.4 ppm) and tetraedrally
(42.6 to 68.0 ppm) coordinated sites.  Although the small occupation of
this site in model C cannot reproduce the $AlO_5$ peak observed by some
experiments, the $\delta_{iso}$ is in the experimentally measured range,
implying that an increase in $AlO_5$ sites due to defects and/or surface
effects would explain the observed feature.

It should be noted that in the study by Pecharrom\'{a}n~{\it et~al.}\cite{PecharromanSobrados1999},
the complex NMR spectrum in the
$\gamma$-alumina region was analyzed as superposition of a small
number of peaks 
for which the quadrupolar interaction parameters were estimated for tetrahedral 
($C_Q$ = 4.7 - 4.9 MHz) and octahedral ($C_Q$ = 3.6 - 3.9 MHz) sites.
These estimated values do not agree with the ones obtained for 
model C.
Since model C shows very good agreement for the total spectrum,
this implies that extracting NMR parameters from a complex spectrum,
in absence of further experimental or theoretical charaterization,
is likely an unreliable procedure as it can result in widely different
distributions of NMR parameters for the same spectrum.


\subsection{Correlation analysis}
\label{CorrelationAnalysis}

\begin{figure*}[htb]
 \includegraphics[width=0.8\textwidth]{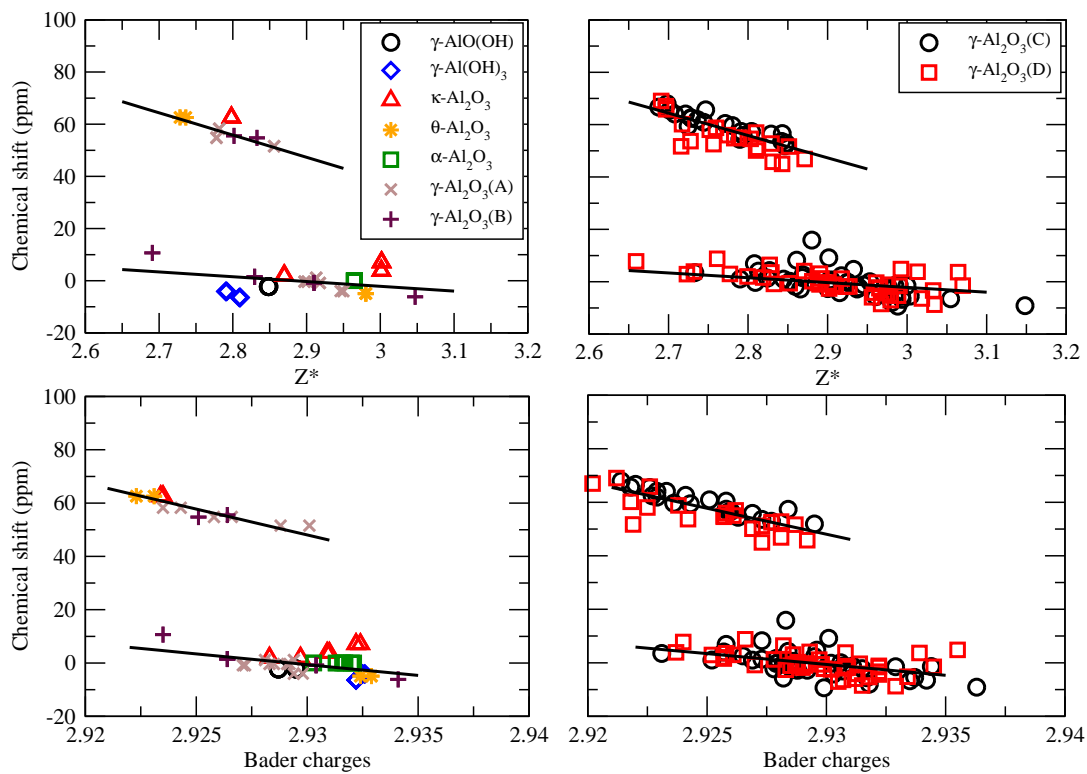}
 \caption{(color on-line) Correlations between chemical shifts and
 effective charges, $Z^*$, (above) or Bader charges (below).  Left panels
 collect the data for all the structures considered in this work except
 the $\gamma$-alumina models C and D, that contain many different $Al$
 sites. For these models the data are reported separately in the right
 panels. Lines correspond to linear fit to $Al_{oct}$ and $Al_{tet}$
 data for all the structures.
}
 \label{fig:correlations}
\end{figure*}

In the literature there have been several attempts to correlate
features in NMR spectra with local structural details such as
coordination numbers~\cite{ChoiMatsunaga2009}, local atomic bondlength and angular
distortion~\cite{MaoWalter2007}, shortest bond-length~\cite{PechkisWalter2011}, Mulliken
charge population~\cite{ChoiMatsunaga2009} etc.

Our calculations confirm the dependence of chemical shifts on aluminium
coordination numbers, such that 4-, 5- and 6-coordinated aluminium sites
show chemical shifts in well separated ranges: the lower the coordination
number the larger is the chemical shift.

However, the internal distribution within a given coordination number
does not correlate with local geometric descriptors such as local
bondlength, angular distortions~\cite{MaoWalter2007}, or shortest
bondlength\cite{PechkisWalter2011}.
To explain this we performed a simple test in the {\it non-spinel}
model B for $\gamma$-alumina by displacing an oxygen atom while keeping
the rest fixed and we observed that the NMR parameters were affected in
a region extending up to the third coordination shell, thus demonstrating
the sensitivity of NMR to non-local structural details.

The only properties that we found to display a significant correlation
with the NMR shieldings were aluminium Bader charges and Born dynamical
effective charges. For both charges the correlation is linear and
structure independent as demostrated by the linear fits obeyed equally well
by all phases, as shown in Fig.~\ref{fig:correlations}.

It should be mentioned that Bader charges and Born effective charges
correlate very strongly with each other so that no additional information
can be gained by considering the two charges together. This is because
$\gamma$-alumina is a crystal with a high degree of ionicity, where
these two quantities are strongly related.

Furthemore, we can understand the correlation of NMR shieldings with
Born effective charges because they both originate from local electronic
susceptibility, one being determined from the current induced by the
magnetic field, the other measuring the charge flow associated to
vibrational motion.

As an hystorical curiosity, we mention that proposing a correlation
between NMR chemical shifts and effective charges is not a complete
novelty since in a couple of papers\cite{Sears-31P-1978,Sears-27Al-1980}
in the late seventies a correlation was reported to exist for binary
semiconductors among Szigeti's effective charges and $^{27}Al$ and
$^{31}P$ NMR chemical shifts.


\section{Conclusions}
\label{conclusions}

We investigated from first principles the $^{27}Al$ NMR properties of
several well characterized crystalline phases of $Al_2O_3$  and of two
of its calcination precursor phases, obtaining very good agreement
with available experimental results. New insight for the peak
assignments in the spectra were proposed for some structures.

This gave confidence in the theoretical approach and allowed us
to address the open problem of the structural characterization of
the technologically important $\gamma$-alumina phase by comparing the
experimental spectra with the theoretical predictions calculated for
four structural models recently appeared in the literature: our study
supports the model structure with $Fd\overline{3}m$ symmetry proposed
by Paglia~{\it et~al.} in Ref.~\on-line-cite{PagliaRohl2005} as the
one that best reproduces the NMR experimental results in the bulk.

Calculations confirm that chemical shifts strongly depend on
coordination number. Moreover, within a given coordination number,
a linear correlation exists between chemical shifts and Born effective
charges or Bader charges.


\section{Acknowledgments}
\label{Acknowledgments}

We are grateful to Dr.\ L.A.\ O'Dell, Dr.\ M.E.\ Smith, and Dr.\ T.J.\
Bastow for providing their experimental NMR spectra and to Dr.\ Gonzalo
Guti\'{e}rrez and Dr.\ Eduardo Men\'{e}ndez-Proupin for providing their
structural model.
SdG and EK like to thank Davide Ceresoli for useful discussions and
for suggesting the use of QuadFit code for the simulation of spectra
including quadrupolar interaction.
ARF wishes to thank SISSA for the support and facilities during the
obtention of the results.
Calculations have been performed on the Sp6-IBM machine at CINECA in Bologna
and on the HPC cluster at SISSA. 
This work was also supported by Petrobras S.A. and brazilian agencies
CAPES, FAPEMIG and CNPq.



\begin{thebibliography}{99}
\bibitem{kmesh}
The k-point meshes for each structure are as follows:
Boehmtie 3x1x3; 
Gibbsite 2x2x2; 
$\alpha$-$Al_2O_3$ 2x2x1;
$\kappa$-$Al_2O_3$ 3x2x2;
$\theta$-$Al_2O_3$ 1x4x2;
$\gamma$-$Al_2O_3$(A) 2x2x1;
$\gamma$-$Al_2O_3$(B) 2x2x1;
$\gamma$-$Al_2O_3$(C) 2x2x1;
$\gamma$-$Al_2O_3$(D) 1x2x1.
All meshes were shifted by half of the mesh interval in each direction.
\end{thebibliography}
\end{document}